\documentclass[12pt,a4paper,dvips]{article}

\usepackage{epsfig}
\usepackage{graphicx}
\usepackage{array}
\usepackage{dcolumn}
\usepackage{multirow}
\usepackage{multicol}
\usepackage{cite}
\mathchardef\d="0064            
\def\diameter{\kern0.3em/\kern-0.72em$\bigcirc$}
\begin{document}

\begin{center}\textbf{\Large {\bf Proton induced damage  in LFS-3 
and LFS-8 scintillating crystals.}}\end{center}

\vspace{1.cm}

\begin{center}
{\large 
V.A. Kozlov$^{1}$,  
S.A. Kutovoi$^{2}$, A.I. Zagumennyi$^{2}$, 
Yu.D. Zavartsev$^{2}$, \\
M.V. Zavertyaev$^{1}$,
A.F. Zerrouk$^{3}$}
\end{center}

\vspace{0.5cm}

{\normalsize 1. P.N. Lebedev
Physical Institute of Russian Academy of Sciences, Moscow, Russia} 
\vspace*{0.5cm}

{\normalsize 2. Prohorov General Physics Institute of Russian Academy of Sciences, 
Moscow, Russia}\vspace*{0.5cm}

{\normalsize 3. Zecotek Imaging Systems Pte Ltd, Division 
of Zecotek Photonics Inc., Vancouver, Canada}

\vspace{1.5cm}

\Large {\bf Abstract}

\bigskip

\normalsize
Scintillating $LFS-3$ and $LFS-8$ crystals were exposed to a 155 MeV/c proton 
fluence
$\Phi_{p}=(4.4\pm0.4)\cdot10^{12}cm^{-2}$. There was negligible reduction in
transmission spectrum of $LFS-3$ crystal measured in 30 days after irradiation.

\section{Introduction}

Construction of the electromagnetic calorimeters  
for the experimental setups at LHC and ILC colliders require use of new radiation 
hard fast materials
such as heavy scintillators and Cherenkov radiators, because   of the planned 
high luminosities for these accelerators.
At present scintillating lead tungstate crystals $PbWO_{4}$ are successfully used 
in ALICE and CMS experiments at 
LHC. It is well known, that dominant radiation damage of scintillating crystals  is 
a hadron induced one because of the large flux  and interaction cross section. 
During last decade 
radiation damages of $PbWO_{4}$ crystals were extensively studied by 
CMS collaboration using pion and proton beams. 
\cite{Huhtinen1,Lecomte1,Lecomte2,Adzic} 
It was found, that high-energy pions and protons cause hadron-specific 
cumulative damage of $PbWO_{4}$ crystals. 
  Studies of CMS collaboration showed, that
an important limitation for the existing lead tungstate crystals at HLLHC 
(and possible even at LHC)
appears to come from ``star'' formation  under irradiation by high energy 
hadrons through nuclear fission.
Therefore, search of heavy, fast and radiation resistant against hadron irradiation  
scintillating crystals is very important now. 
Recently CMS collaboration has published their results on proton induced damage 
in scintillating $CeF_{3}$ crystals,
a potential candidate  for electromagnetic calorimetry at HLLHC. \cite{Dissertori}

 Below we present the first results on measurements of radiation damage 
of $LFS$ crystals  caused by 155 MeV/c proton beam.

\section{Results and discussion}

\begin{table}[htp]
\caption{The basic properties of the scintillating crystals.}
\label{tab:prop}
\begin{center}
\begin{tabular}{|c|c|c|c|} \hline
\normalsize Material & \normalsize $NaI(Tl)$ 
&\normalsize LFS-3 
& \normalsize LFS-8
\\ \hline\hline
\normalsize Density, $\rho$ $(g/cm^{3})$   &\normalsize   3.67 
&\normalsize   7.35 
&\normalsize   7.4
\\ \hline

\normalsize Melting point, $(^{0}C)$       &\normalsize   651  
&\normalsize   2000 
&\normalsize   2000
\\\hline

\normalsize Radiation length, $X_{0}$ (cm) &\normalsize   2.59 
&\normalsize   1.15 
&\normalsize   1.14
\\\hline

\normalsize Moliere radius, $R_{m}$ (cm)   &\normalsize   4.3  
& \normalsize  2.09 
& \normalsize  2.07
\\\hline

\normalsize Light output ($\%$)            &\normalsize   100  
& \normalsize   85  
& \normalsize   82
\\\hline

\normalsize Decay time, (ns)               &\normalsize   230  
&\normalsize    35  
&\normalsize    19
\\\hline

\normalsize Peak emission, (nm)            &\normalsize   410  
& \normalsize  425  
& \normalsize  422
\\\hline

\normalsize Refractive index, $n$          &\normalsize   1.85 
& \normalsize  1.81 
& \normalsize  1.81
\\
\normalsize in maximum of emission         &              
&              
&              
\\\hline

\normalsize Hardness, (Moh)                &\normalsize    2   
&\normalsize    5   
&\normalsize    5
\\\hline

\normalsize Hygroscopic                    &\normalsize   Yes  
&\normalsize   No   
&\normalsize   No
\\\hline

\end{tabular} 
\end{center}
\end{table}

Proprietary, bright scintillators $LFS$ (Lutetium Fine Silicate)
developed by
Zecotek Imaging Systems Pte Ltd provide much improved scintillating 
parameters and reproducibility \cite{Zagumennyi}.
$LFS$ is a brand name of the set of Ce-doped scintillation crystals of the solid
solutions on the basis of the silicate crystal, comprising lutetium and
crystallizing in the
monoclinic system, spatial group $C2/c$, $Z=4$. The patented $LFS$
compositions is $Ce_{x}Lu_{2+2y-x-z}
A_{z}Si_{1-y}O_{5+y}$, where A is at least one
element selected from
the group consisting of $Ca$, $Gd$, $Sc$, $Y$, $La$, $Eu$ and $Tb$.
The raw materials were $99.999\%$ pure $Lu_{2}O_{3}$, $SiO_{2}$ and the 
scintillating $CeO_{2}$ dopant.
 
The $LFS$ crystals demonstrated stable scintillation parameters for top 
and bottom ends of large boules in
comparison with $LSO$. The most important parameters of two types of
 $LFS$ scintillating crystals 
in comparison with characteristics of common inorganic
scintillator $NaI(Tl)$ are presented in Table\,\ref{tab:prop}. The main
properties of $LFS$ crystal make it highly suitable as a scintillating 
material for electromagnetic
calorimeters in high energy particle physics experiments.
Earlier radiation damage of new heavy $LFS-3$ crystal has been studied
using powerful $^{60}Co$ source at the dose rate of 4 Krad/min.
No deterioration in optical transmission of $LFS-3$ crystal was observed
after irradiation with the dose 23 Mrad. \cite{Kozlov}

The large $LFS-3$ and $LFS-8$ crystals were grown by Zecotek Imaging Systems 
Pte Ltd, Division of
Zecotek Photonics Inc., Vancouver, Canada with the Czochralski technique. 
The initial crystal boules have been cut up to the samples with the 
dimensions of $11\times11$ $mm^{2}$ and 20 mm long. All crystals
samples have been polished to an optical grade. 
The crystals were packed to $3\times2$ matrix for simultaneous irradiation 
with proton beam from ITEP proton synchrotron. The proton beam with diameter 
$\sim$ 50 mm was parallel to long size of $3\times2$ crystal matrix. The beam 
uniformity was about $\le 5\%$ over the whole beam spot. All crystals have been 
irradiated to a 155 MeV/c protons up to fluence of $4.4\cdot10^{12}$ p/cm$^{2}$.
 Optical transmission spectra across a 20 mm thickness
were measured with a spectrophotometer (Shimadzu UV-3101PC)
 before and at various intervals after proton irradiation.

\begin{figure}[htb] 
\addtolength{\abovecaptionskip}{6pt}
\begin{minipage}[l]{8.cm}
\centering
\includegraphics[width=8cm]{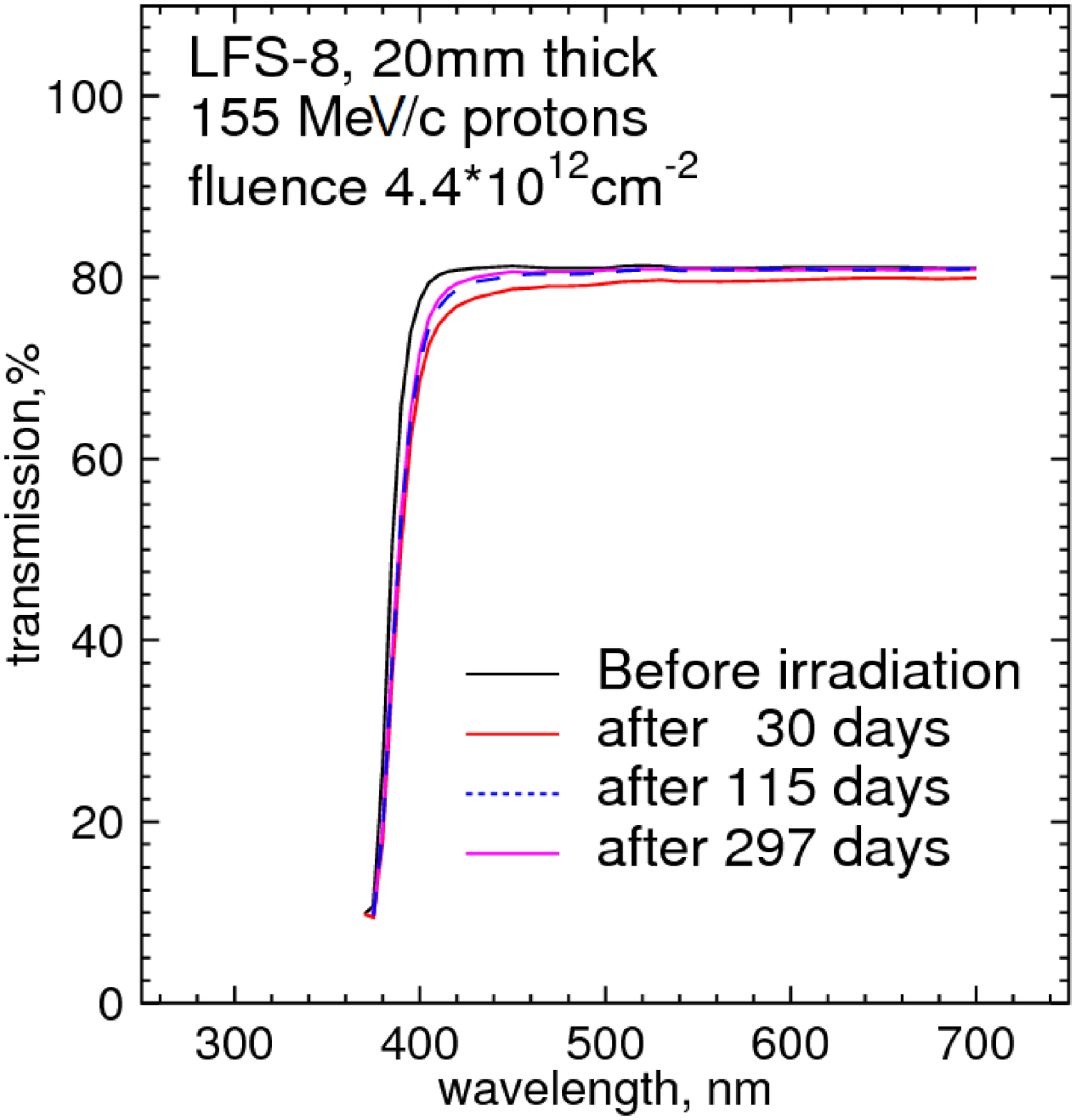}
\parbox{7.5cm}{\caption{\normalsize \ \ \ \ \ Transmission spectra of \mbox
{$LFS-8$} crystal before and 
at various intervals after irradiation(sample thickness 20 mm).}} 
\label{fig:trans1}
\end{minipage}
\begin{minipage}[r]{8.cm}
\centering
\includegraphics[width=8cm]{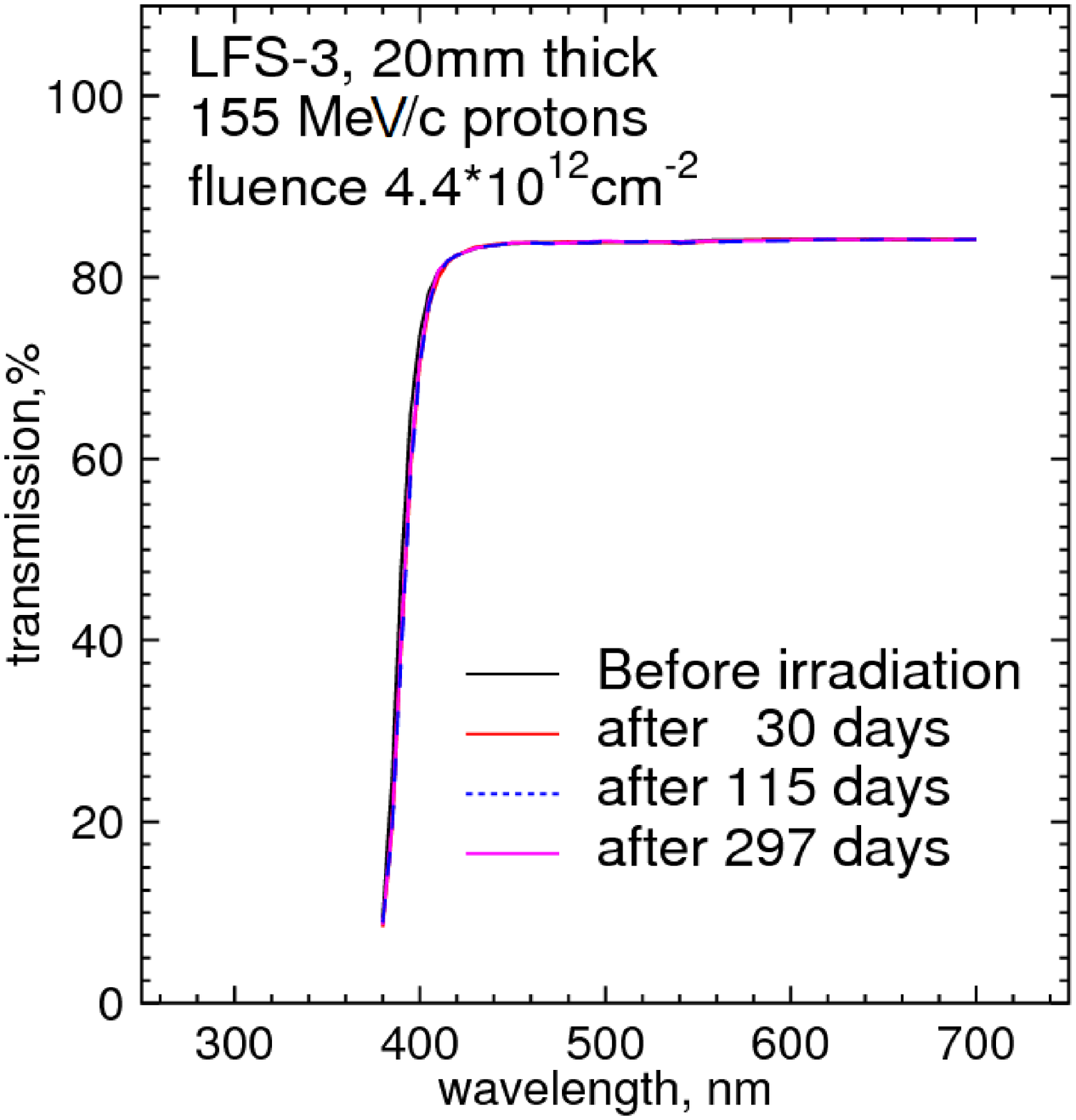}
\parbox{7.5cm}{\caption{\normalsize \ \ \ \ \ Transmission spectra of \mbox
{$LFS-3$} crystal before and 
at various intervals after proton irradiation(sample thickness 20 mm).}} 
\label{fig:trans2}
\end{minipage}
\end{figure} 

 Due to induced radioactivity of $LFS$ crystals first measurements of optical transmission 
of crystals were made in 30 days after proton irradiation.

The transmission spectra for $LFS-8$ 
 and $LFS-3$ 
 crystals are presented in 
Fig.\,1 and Fig.\,2.  Spontaneous recovery at room temperature was 
observed for $LFS-8$ crystal during ten months after irradiation.
 It is evident, that irradiated $LFS-3$ crystal has negligible reduction
in transmission spectrum measured in 30 days after irradiation. 



\section {Conclusions}
 The results of studies on radiation damage of different $LFS$ crystals by 
using 155 MeV/c proton beam from ITEP proton synchrotron are presented.
A significant progress has been made in the development of $LFS$ crystals 
which could be used as the active medium of a fast performance electromagnetic 
calorimeter  for a wide range of particle physics application.
The obtained results indicate, that $LFS-3$ seems to be the best in terms  
of radiation resistance against hadron irradiation among all other
crystals used in high-energy physics experiments. 


\end{document}